\documentclass[11pt]{article}
\usepackage{moriond,epsfig}
\usepackage{units,amssymb}

\newcommand{\dm}{\partial_\mu}

\begin{document}
\vspace*{4cm}
\title{The Standard model Higgs as the inflaton}

\author{F. L. Bezrukov}

\address{
  Institut de Th\'eorie des Ph\'enom\`enes Physiques,
  \'Ecole Polytechnique F\'ed\'erale de Lausanne,
  CH-1015 Lausanne, Switzerland}

\maketitle
\abstracts{%
  We describe how non-minimal coupling term between the Higgs boson
  and gravity can lead to the chaotic inflation in the Standard Model
  without introduction of any additional degrees of freedom.  Produced
  cosmological perturbations are predicted to be in accordance with
  observations.  The tensor modes of perturbations are practically
  vanishing in the model.}

\section{Introduction}

This talk is based on the recent work \cite{Bezrukov:2007ep}, and
closely follows it.  Note, that the expression for the inflationary
potential presented here differs from the one presented in the
original work---both expressions coincide in the region relevant for
inflation, while the expression given here has a wider range of
validity (down to the Standard Model regime).

The fact that our universe is almost flat, homogeneous and isotropic
is often considered as a strong indication that the Standard Model
(SM) of elementary particles is not complete.  Indeed, these puzzles,
together with the problem of generation of (almost) scale invariant
spectrum of perturbations, necessary for structure formation, are most
elegantly solved by inflation~%
\cite{Starobinsky:1979ty,Starobinsky:1980te,Mukhanov:1981xt,Guth:1980zm,%
  Linde:1981mu,Albrecht:1982wi}.  The majority of present models of
inflation require an introduction of an additional scalar---the
``inflaton''.  Inflaton properties are constrained by the observations
of fluctuations of the Cosmic Microwave Background (CMB) and the
matter distribution in the universe.  Though the mass and the
interaction of the inflaton with matter fields are not fixed, the well
known considerations prefer a heavy scalar field with a mass $\sim
\unit[10^{13}]{GeV}$
and extremely small self-interacting quartic
coupling constant $\lambda \sim 10^{-13}$ for realization of the
chaotic inflationary scenario~\cite{Linde:1983gd}.  This value of the
mass is close to the GUT scale, which is often considered as an
argument in favour of existence of new physics between the electroweak
and Planck scales.

It was recently demonstrated in \cite{Bezrukov:2007ep} that the SM
itself can give rise to inflation, provided non-minimal copling of the
Higgs field with gravity.  The spectral index and the amplitude of
tensor perturbations can be predicted and be used to distinguish this
possibility from other models for inflation; these parameters for the
SM fall within the $1\sigma$ confidence contours of the WMAP-5
observations \cite{Komatsu:2008hk}.

To explain our main idea, let us consider the Lagrangian of
the SM non-minimally coupled to gravity,
\begin{equation}
\label{main}
  L_{\mathrm{tot}}= L_{\mathrm{SM}} - \frac{M^2}{2} R -\xi H^\dagger HR
  \;,
\end{equation}
where $L_{\mathrm{SM}}$ is the SM part, $M$ is some mass parameter,
$R$ is the scalar curvature, $H$ is the Higgs field, and $\xi$ is an
unknown constant to be fixed later.  The third term in (\ref{main}) is
in fact required by the renormalization properties of the scalar field
in a curved space-time background \cite{Birrell:1982ix}, so, in
principle, it should be added to the usual SM Lagrangian with some
constant.  Here, we will analyse the situation with large
non-minimal coupling parameter $\xi\gg1$, but still not too large for
the non-minimal term to contribute significantly to the Plank mass in
the SM regime ($H\sim v$), i.e.\ $\sqrt{\xi} \lll 10^{17}$.  Thus, we
have $M\simeq M_P=(8\pi G_N)^{-1/2}=\unit[2.4\times 10^{18}]{GeV}$.

It is well known that inflation has interesting properties in models
of this type~%
\cite{Spokoiny:1984bd,Futamase:1987ua,Salopek:1988qh,Fakir1990,Kaiser:1994wj,%
  Kaiser:1994vs,Komatsu:1999mt}.  However, in these works the scalar
was not identified with the Higgs field of the SM.  Basically, most
attempts were made to identify the inflaton field with the GUT Higgs
field.  In this case one naturally gets into the regime of induced
gravity (where, unlike this paper, $M=0$ and $M_P$ is generated from
the non-minimal coupling term by the Higgs vacuum expectation value).
In this case the Higgs field decouples from the other fields of the
model \cite{vanderBij:1993hx,CervantesCota:1995tz,Bij1995}, which is
generally undesirable.  Here we demonstrate, that when the SM Higgs
boson is coupled non-minimally to gravity, the scales for the
electroweak physics and inflation are separate, the electroweak
properties are unchanged, while for much larger field values the
inflation is possible.

The paper is organised as follows.  We start from discussion of
inflation in the model, and use the slow-roll approximation to find
the perturbation spectra parameters.  Then we will argue in
Section~\ref{sec:radcorr} that quantum corrections are unlikely to
spoil the classical analysis we used in Section~\ref{sec:cmb}.  We
conclude in Section~\ref{sec:concl}.

\section{Inflation and CMB fluctuations}
\label{sec:cmb}

Let us consider the scalar sector of the Standard Model, coupled to
gravity in a non-minimal way. We will use the unitary gauge
$H=h/\sqrt{2}$ and neglect all gauge interactions for the time being,
they will be discussed later in Section \ref{sec:radcorr}.  Then the
Lagrangian has the form:
\begin{equation}
  \label{eq:1}
    S_{J} =\int d^4x \sqrt{-g} \Bigg\{
    - \frac{M^2+\xi h^2}{2}R
    \\
    + \frac{\dm h\partial^\mu h}{2}
    -\frac{\lambda}{4}\left(h^2-v^2\right)^2
    \Bigg\}
    \;.
\end{equation}
This Lagrangian has been studied in detail in many papers on inflation
\cite{Salopek:1988qh,Fakir1990,Kaiser:1994vs,Komatsu:1999mt}, we will
reproduce here the main results of
\cite{Salopek:1988qh,Kaiser:1994vs}.  Compared to
\cite{Bezrukov:2007ep} we present a better approximation for the
inflationary potential here.  To simplify the formulae, we will
consider only $\xi$ in the region $1\ll\sqrt{\xi}\lll10^{17}$, in
which $M \simeq M_P$ with very good accuracy.

\begin{figure}
  \centering
  \begin{minipage}[t]{0.5\textwidth}
    \includegraphics[width=\textwidth]{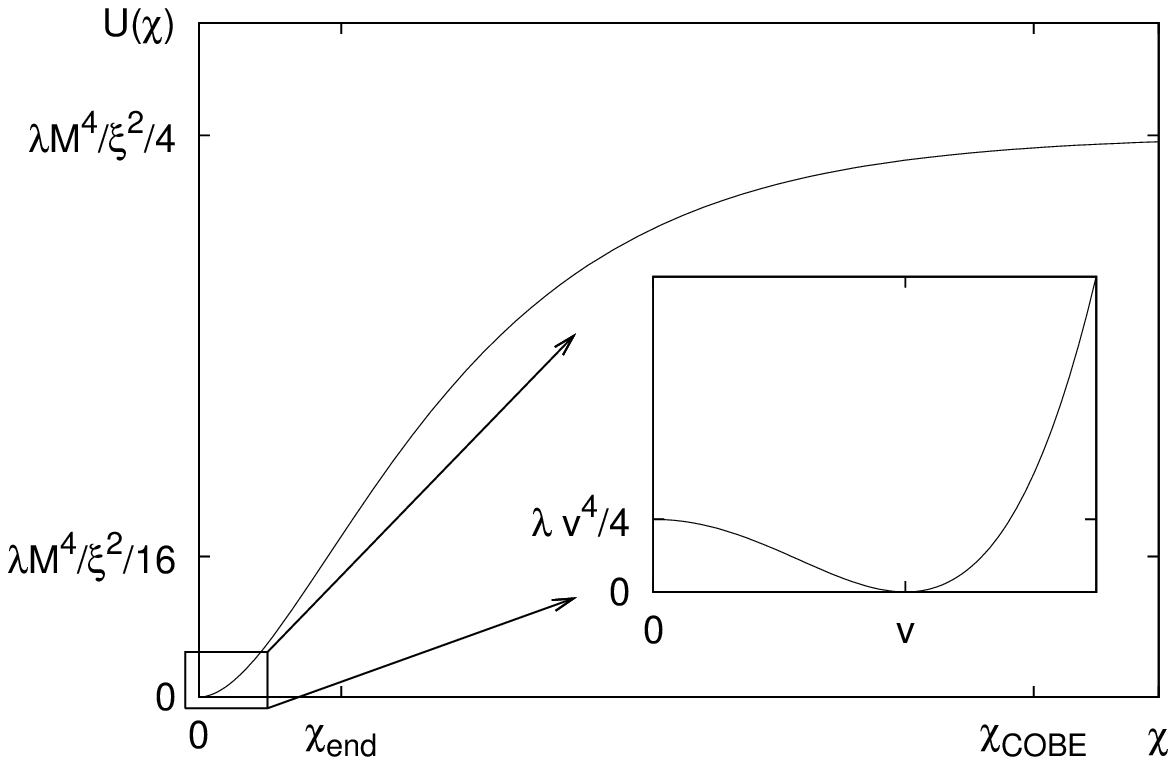}
    \caption{Effective potential in the Einstein frame.}
    \label{fig:Ueff}
  \end{minipage}%
  \begin{minipage}[t]{0.5\textwidth}
    \includegraphics[width=0.865\textwidth]{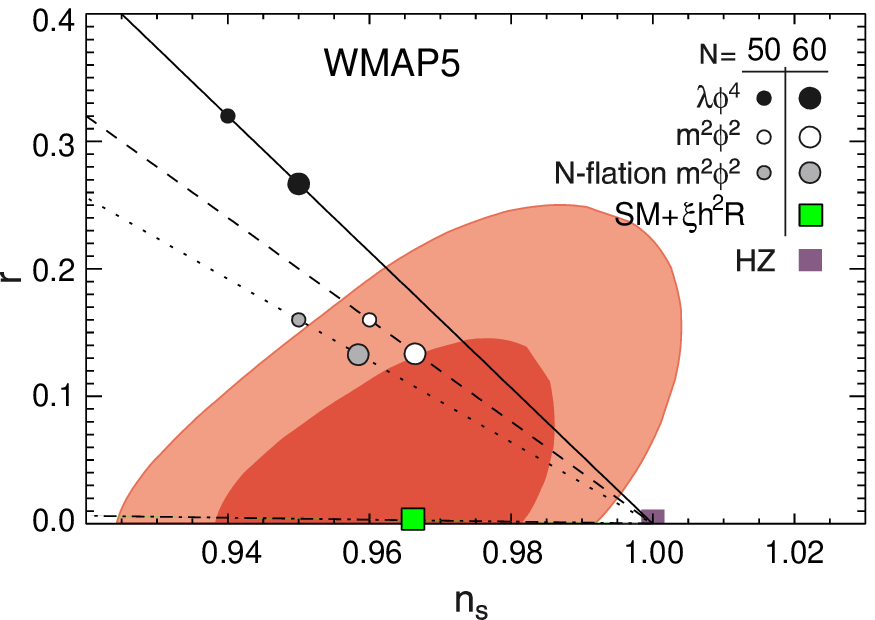}
    \caption{The allowed WMAP region for inflationary parameters ($r$,
      $n$).  The green boxes are our predictions supposing 50 and 60
      e-foldings of inflation.  Black and white dots are predictions of
      usual chaotic inflation with $\lambda\phi^4$ and $m^2\phi^2$
      potentials, HZ is the Harrison-Zeldovich spectrum.}
    \label{fig:wmap}
  \end{minipage}
\end{figure}

It is possible to get rid of the non-minimal coupling to gravity by
making the conformal transformation from the Jordan frame to the
Einstein frame
\begin{equation}
  \label{eq:2}
  \hat{g}_{\mu\nu} = \Omega^2 g_{\mu\nu}
  \;,\quad
  \Omega(h)^2 = 1 + \frac{\xi h^2}{M_P^2}
  \;.
\end{equation}
This transformation leads to a non-minimal kinetic term for the Higgs
field. So, it is convenient to make the change to the new scalar field
$\chi$ with
\begin{equation}
  \label{eq:3}
   \frac{d\chi}{dh}=\frac{\sqrt{\Omega^2+\frac{3}{2}M_P^2\left(\frac{d(\Omega^2)}{dh}\right)^2}}{\Omega^2}
   =\frac{\sqrt{1 + (\xi+6\xi^2)\frac{h^2}{M_P^2}}}{1 + \xi\frac{ h^2}{M_P^2}}
  \;.
\end{equation}
Finally, the action in the Einstein frame is
\begin{equation}
  \label{eq:4}
    S_E =\int d^4x\sqrt{-\hat{g}} \Bigg\{
    - \frac{M_P^2}{2}\hat{R}
    + \frac{\dm \chi\partial^\mu \chi}{2}
    - U(\chi)
    \Bigg\}
    \;,
\end{equation}
where $\hat{R}$ is calculated using the metric $\hat{g}_{\mu\nu}$ and
the potential is
\begin{equation}
  \label{eq:5}
  U(\chi) =
  \frac{1}{\Omega(h(\chi))^4}\frac{\lambda}{4}\left(h(\chi)^2-v^2\right)^2
  \;.
\end{equation}
For small field values $h,\chi<M_P/\xi$ the change of variables is
trivial, $h\simeq\xi$ and $\Omega^2\simeq1$, so the potential for the
field $\chi$ is the same as that for the initial Higgs field and we
get into the SM regime.  For $h,\chi\gg M_P/\xi$ the situation changes
a lot.  In this limit the variable change (\ref{eq:3}) is
\footnote{The following two formulae have wider validity range than
  those in \cite{Bezrukov:2007ep}, which are valid only for
  $h\gg M_P/\sqrt{\xi}$.}
\begin{equation}\label{eq:hlarge}
  \Omega(h)^2\simeq \exp\left(\frac{2\chi}{\sqrt{6}M_P}\right)
  \;.
\end{equation}
The potential for the Higgs field is exponentially flat for large
$\xi$ and has the form
\begin{equation}
  \label{eq:6}
  U(\chi) = \frac{\lambda M_P^4}{4\xi^2}
  \left(
    1-\exp\left(
      -\frac{2\chi}{\sqrt{6}M_P}
    \right)
  \right)^{2}
  \;.
\end{equation}
The full effective potential in the Einstein frame is presented in
Fig.~\ref{fig:Ueff}.  It is the flatness of the potential at
$\chi\gtrsim M_P$ which makes the successful (chaotic) inflation
possible.

Basically, there are two distinct scales---for low field values
$h,\chi\ll
M_P/\xi$ we have the SM, for high field values
$h\gg M_P/\sqrt{\xi}$ ($\chi>M_P$) we have inflation with exponentially flat
potential (\ref{eq:6})
and the Higgs field is decoupled from all other SM fields (because $\Omega\propto
h$,
see Section~\ref{sec:radcorr}).  In the intermediate region
$M_P/\xi\ll h\ll M_P/\sqrt{\xi}$ ($M_P/\xi\ll\chi<M_P$) the coupling with other particles is
not suppressed ($\Omega\sim 1$), while the potential and change of
variables are still given by (\ref{eq:6}) and (\ref{eq:hlarge}).

Analysis of the inflation in the Einstein frame \footnote{The same
  results can be obtained in the Jordan frame
  \cite{Makino1991,Fakir:1992cg}.} can be performed in the standard way
using the slow-roll approximation.  The slow roll parameters (in
notations of \cite{Lyth:1998xn}) can be expressed analytically as
functions of the field $h(\chi)$ using (\ref{eq:3}) and (\ref{eq:5})
(we give here the expressions for the case
\footnote{These formulas are valid up to the end of the slow roll regime
  $h_\mathrm{end}$, while the formulas (10) and (11) in
  \cite{Bezrukov:2007ep} are applicable only for the earlier
  inflationary stages, $h^2\gg M_P^2/\xi$, which is sufficient to
  calculate primordial spectrum parameters $n_s$ and $r$.}
$h^2\gtrsim M_P^2/\xi\gg v^2$, $\xi\gg1$, exact
expressions can be found in \cite{Kaiser:1994vs}),
\begin{eqnarray}
  \label{eq:7}
  \epsilon & =& \frac{M_P^2}{2}\left(\frac{dU/d\chi}{U}\right)^2
  \simeq\frac{4 M_P^4 }{3
   \xi^2h^4}
  \;, \\
  \eta & = & M_P^2\frac{d^2U/d\chi^2}{U}
  \simeq \frac{4 M_P^4}{3 \xi^2 h^4 }\left(1-\frac{\xi h^2}{M_P^2}\right)
  \;, \\
  \zeta^2 &= & M_P^4\frac{(d^3U/d\chi^3)dU/d\chi}{U^2}
  \simeq \frac{16 M_P^6 }{9\xi^3 h^6}\left(\frac{\xi h^2}{M_P^2}-3\right)
  \;.
\end{eqnarray}
Slow roll ends when $\epsilon\simeq1$, so the field value at the end of
inflation is
$h_{\mathrm{end}}\simeq(4/3)^{1/4}M_P/\sqrt{\xi}\simeq1.07M_P/\sqrt{\xi}$.
The number of e-foldings for the change of the field $h$ from $h_0$ to
$h_{\mathrm{end}}$ is given by
\begin{equation}
  \label{eq:8}
  N = \int_{h_{\mathrm{end}}}^{h_0}
  \frac{1}{M_P^2}\frac{U}{dU/dh}\left(\frac{d\chi}{dh}\right)^2dh
  \simeq \frac{3}{4}\frac{h_0^2-h_{\mathrm{end}}^2}{M_P^2/\xi}
  \;.
\end{equation}
We see that for all values of $\sqrt{\xi}\lll10^{17}$ the scale
of the Standard Model $v$ does not enter in the formulae, so the
inflationary physics is independent on it.

After end of the slow roll the $\chi$ field enters oscillatory stage
with diminishing amplitude.  After the oscillation amplitude falls
below $M_P/\xi$, the situation returns to the SM one, so at this
moment the reheating is imminent due to the SM interactions, which
guarantees the minimum reheating temperature
$T_{\mathrm{reh}}\gtrsim
(\frac{15\lambda}{8\pi^2 g^*})^{1/4}\frac{M_P}{\xi}\simeq\unit[1.5\times
10^{13}]{GeV}$,
where $g^*=106.75$ is the number of degrees of freedom
of the SM\@.  Careful analysis may give a larger temperature generated
during the decay of the oscillating $\chi$ field, but definitely below
the energy scale at the end of the inflation
$T_{\mathrm{reh}}<(\frac{2\lambda}{\pi^2
  g^*})^{1/4}\frac{M_P}{\sqrt{\xi}}\simeq\unit[2\times10^{15}]{GeV}$.

As far as the reheating mechanism and the universe evolution after the
end of the inflation is fixed in the model, the number of e-foldings
for the the COBE scale entering the horizon can be calculated (see
\cite{Lyth:1998xn}).  Here we estimate it as
$N_{\mathrm{COBE}}\simeq62$ (exact value depends on the detailed
analysis of reheating, which will be done elsewhere).  The
corresponding field value is
$h_{\mathrm{COBE}}\simeq9.4M_P/\sqrt{\xi}$.  Inserting (\ref{eq:8})
into the COBE normalization $U/\epsilon=(0.027M_P)^4$ we find the
required value for $\xi$
\begin{equation}
  \label{eq:9}
  \xi \simeq \sqrt{\frac{\lambda}{3}}\frac{N_{\mathrm{COBE}}}{0.027^2}
  \simeq  49000\sqrt{\lambda}
  =  49000\frac{m_H}{\sqrt{2}v}
  \;.
\end{equation}
Note, that if one could deduce $\xi$ from some fundamental theory this
relation would provide a connection between the Higgs mass and the
amplitude of primordial perturbations.

The spectral index $n_s=1-6\epsilon+2\eta$ calculated for $N=60$
(corresponding to the scale $k=0.002/\mathrm{Mpc}$) is
$n_s\simeq1-8(4N+9)/(4N+3)^2\simeq0.97$.  The tensor to scalar
perturbation ratio \cite{Komatsu:2008hk} is
$r=16\epsilon\simeq192/(4N+3)^2\simeq0.0033$.
The predicted values are well within one sigma of the current WMAP
measurements \cite{Komatsu:2008hk}, see Fig.~\ref{fig:wmap}.

\section{Radiative corrections}
\label{sec:radcorr}

An essential point for inflation is the flatness of the scalar
potential in the region of the field values $h\sim10M_P/\sqrt{\xi}$
($\chi\sim 6 M_P$).  It is important that radiative corrections do not
spoil this property.  Of course, any discussion of quantum corrections
is flawed by the non-renormalizable character of gravity, so the
arguments we present below are not rigorous.

There are two qualitatively different type of corrections one can
think about.  The first one is related to the quantum gravity
contribution.  It is conceivable to think \cite{Linde:1987yb} that
these terms are proportional to the energy density of the field $\chi$
rather than its value and are of the order of magnitude $U(\chi)/M_P^4
\sim \lambda/\xi^2$.
They are small at large $\xi$ required by
observations.  Moreover, adding non-renormalizable operators
$h^{4+2n}/M_P^{2n}$ to the Lagrangian (\ref{eq:1}) also does not
change the flatness of the potential in the inflationary
region.\footnote{Actually, in the Jordan frame, we expect that
  higher-dimensional operators are suppressed by the effective Planck
  scale $M_P^2+\xi h^2$.}

Other type of corrections is induced by the fields of the Standard
Model coupled to the Higgs field.  In one loop approximation these
contributions have the structure
\begin{equation}
\Delta U \sim \frac{m^4(\chi)}{64\pi^2} \log\frac{m^2(\chi)}{\mu^2}~,
\label{1loop}
\end{equation}
where $m(\chi)$ is the mass of the particle (vector boson, fermion, or
the Higgs field itself) in the background of field $\chi$, and $\mu$ is
the normalization point.  Note that the terms of the type $m^2(\chi)
M_P^2$
(related to quadratic divergences) do not appear in
scale-invariant subtraction schemes that are based, for example, on
dimensional regularisation (see a relevant discussion in
\cite{Shaposhnikov:2006xi,Meissner:2006zh,Shaposhnikov:2007nj,Meissner:2007xv}).
The masses of the SM fields can be readily computed
\cite{Salopek:1988qh} and have the form
\begin{equation}
  m_{\psi,A}(\chi) = \frac{m(v)}{v}\frac{h(\chi)}{\Omega(\chi)}
  \;,\quad
  m^2_H(\chi) = \frac{d^2U}{d\chi^2}
\end{equation}
for fermions, vector bosons and 
the Higgs (inflaton) field.  It is crucial that for large $\chi$
these masses approach different constants (i.e.\ the one-loop contribution
is as flat as the tree potential) and that (\ref{1loop}) is suppressed
by the gauge or Yukawa couplings in comparison with the tree term.  In
other words, one-loop radiative corrections do not spoil the flatness
of the potential as well.  This argument is identical to the one given
in \cite{Salopek:1988qh}.



\section{Conclusions}
\label{sec:concl}

Non-minimal coupling of the Higgs field to gravity leads to the
possibility of chaotic inflation in SM.  Specific predictions for the
primordial perturbation spectrum are obtained.  Specifically, very
small amount of tensor perturbations is expected, which means that
future CMB experiments measuring the B-mode of the CMB polarization
(PLANCK) can distinguish between the described scenario from other
models (based, e.g.\ on inflaton with quadratic potential).

At the same time, we expect that the Higgs potential does not enter
into the string coupling regime, nor generates another vacuum up to
the scale of at least $M_P/\xi\sim\unit[10^{14}]{GeV}$, so we expect the
Higgs mass to be in the window $\unit[130]{GeV}<M_H<\unit[190]{GeV}$
(see, eg.\ \cite{Pirogov:1998tj}), otherwise the inflation would be
impossible.

The inflation mechanism we discussed has in fact a general character
and can be used in many extensions of the SM. Thus, the $\nu$MSM of
\cite{Asaka:2005an,Asaka:2005pn,Bezrukov:2005mx,Asaka:2006ek,Shaposhnikov:2006xi,%
  Shaposhnikov:2006nn,Asaka:2006rw,Asaka:2006nq,Bezrukov:2006cy,Gorbunov:2007ak,%
  Shaposhnikov:2008pf,Laine:2008pg} (SM plus three light fermionic
singlets) can explain simultaneously neutrino masses, dark matter,
baryon asymmetry of the universe and inflation without introducing any
additional particles (the $\nu$MSM with the inflaton was considered in
\cite{Shaposhnikov:2006xi}).  This provides an extra argument in favour
of absence of a new energy scale between the electroweak and Planck
scales, advocated in \cite{Shaposhnikov:2007nj}.

\section*{Acknowledgements}

The author thank M. Shaposhnikov, S. Sibiryakov, V. Rubakov, G.
Dvali, I. Tkachev, O.  Ruchayskiy, H.D. Kim, P. Tinyakov, and A.
Boyarsky for valuable discussions.  This work was supported by the
Swiss National Science Foundation.

\section*{References}

\begin{thebibliography}{10}

\bibitem{Bezrukov:2007ep}
F.L. Bezrukov and M. Shaposhnikov,
\newblock Phys. Lett. B659 (2008) 703.

\bibitem{Starobinsky:1979ty}
A.A. Starobinsky,
\newblock JETP Lett. 30 (1979) 682.

\bibitem{Starobinsky:1980te}
A.A. Starobinsky,
\newblock Phys. Lett. B91 (1980) 99.

\bibitem{Mukhanov:1981xt}
V.F. Mukhanov and G.V. Chibisov,
\newblock JETP Lett. 33 (1981) 532.

\bibitem{Guth:1980zm}
A.H. Guth,
\newblock Phys. Rev. D23 (1981) 347.

\bibitem{Linde:1981mu}
A.D. Linde,
\newblock Phys. Lett. B108 (1982) 389.

\bibitem{Albrecht:1982wi}
A. Albrecht and P.J. Steinhardt,
\newblock Phys. Rev. Lett. 48 (1982) 1220.

\bibitem{Linde:1983gd}
A.D. Linde,
\newblock Phys. Lett. B129 (1983) 177.

\bibitem{Komatsu:2008hk}
WMAP, E. Komatsu et~al.,
\newblock (2008), arXiv:0803.0547 [astro-ph].

\bibitem{Birrell:1982ix}
N.D. Birrell and P.C.W. Davies,
\newblock Quantum Fields in Curved Space (Cambridge, UK: Univ. Pr., 1982).

\bibitem{Spokoiny:1984bd}
B.L. Spokoiny,
\newblock Phys. Lett. B147 (1984) 39.

\bibitem{Futamase:1987ua}
T. Futamase and K. Maeda,
\newblock Phys. Rev. D39 (1989) 399.

\bibitem{Salopek:1988qh}
D.S. Salopek, J.R. Bond and J.M. Bardeen,
\newblock Phys. Rev. D40 (1989) 1753.

\bibitem{Fakir1990}
R. Fakir and W.G. Unruh,
\newblock Phys. Rev. D41 (1990) 1783.

\bibitem{Kaiser:1994wj}
D.I. Kaiser,
\newblock Phys. Lett. B340 (1994) 23.

\bibitem{Kaiser:1994vs}
D.I. Kaiser,
\newblock Phys. Rev. D52 (1995) 4295.

\bibitem{Komatsu:1999mt}
E. Komatsu and T. Futamase,
\newblock Phys. Rev. D59 (1999) 064029.

\bibitem{vanderBij:1993hx}
J.J. van~der Bij,
\newblock Acta Phys. Polon. B25 (1994) 827.

\bibitem{CervantesCota:1995tz}
J.L. Cervantes-Cota and H. Dehnen,
\newblock Nucl. Phys. B442 (1995) 391.

\bibitem{Bij1995}
J.J. van~der Bij,
\newblock Int.J.Phys. 1 (1995) 63.

\bibitem{Makino1991}
N. Makino and M. Sasaki,
\newblock Prog. Theor. Phys. 86 (1991) 103.

\bibitem{Fakir:1992cg}
R. Fakir, S. Habib and W. Unruh,
\newblock Astrophys. J. 394 (1992) 396.

\bibitem{Lyth:1998xn}
D.H. Lyth and A. Riotto,
\newblock Phys. Rept. 314 (1999) 1.

\bibitem{Linde:1987yb}
A.D. Linde,
\newblock Phys. Lett. B202 (1988) 194.

\bibitem{Shaposhnikov:2006xi}
M. Shaposhnikov and I. Tkachev,
\newblock Phys. Lett. B639 (2006) 414.

\bibitem{Meissner:2006zh}
K.A. Meissner and H. Nicolai,
\newblock Phys. Lett. B648 (2007) 312.

\bibitem{Shaposhnikov:2007nj}
M. Shaposhnikov,
\newblock (2007), arXiv:0708.3550 [hep-th].

\bibitem{Meissner:2007xv}
K.A. Meissner and H. Nicolai,
\newblock (2007), arXiv:0710.2840 [hep-th].

\bibitem{Pirogov:1998tj}
Y.F. Pirogov and O.V. Zenin,
\newblock Eur. Phys. J. C10 (1999) 629.

\bibitem{Asaka:2005an}
T. Asaka, S. Blanchet and M. Shaposhnikov,
\newblock Phys. Lett. B631 (2005) 151.

\bibitem{Asaka:2005pn}
T. Asaka and M. Shaposhnikov,
\newblock Phys. Lett. B620 (2005) 17.

\bibitem{Bezrukov:2005mx}
F. Bezrukov,
\newblock Phys. Rev. D72 (2005) 071303.

\bibitem{Asaka:2006ek}
T. Asaka, M. Shaposhnikov and A. Kusenko,
\newblock Phys. Lett. B638 (2006) 401.

\bibitem{Shaposhnikov:2006nn}
M. Shaposhnikov,
\newblock Nucl. Phys. B763 (2007) 49.

\bibitem{Asaka:2006rw}
T. Asaka, M. Laine and M. Shaposhnikov,
\newblock JHEP 06 (2006) 053.

\bibitem{Asaka:2006nq}
T. Asaka, M. Laine and M. Shaposhnikov,
\newblock JHEP 0701 (2007) 091.

\bibitem{Bezrukov:2006cy}
F. Bezrukov and M. Shaposhnikov,
\newblock Phys. Rev. D75 (2007) 053005.

\bibitem{Gorbunov:2007ak}
D. Gorbunov and M. Shaposhnikov,
\newblock (2007), arXiv:0705.1729 [hep-ph].

\bibitem{Shaposhnikov:2008pf}
M. Shaposhnikov,
\newblock (2008), arXiv:0804.4542 [hep-ph].

\bibitem{Laine:2008pg}
M. Laine and M. Shaposhnikov,
\newblock (2008), arXiv:0804.4543 [hep-ph].

\end{thebibliography}

\end{document}